\input amstex
\documentstyle{amsppt}
\def\const{\operatorname{const}}
\topmatter
\title
On the point transformations for the second order
differential equations. \uppercase\expandafter{\romannumeral 1}.
\endtitle
\rightheadtext{On the point transformations \dots}
\author
V.V.~Dmitrieva and R.A.~Sharipov
\endauthor
\thanks
Paper is written under the financial support of INTAS foundation
(project \#93-47, coordinator S.I.~Pinchuk) and Russian Fund
for Fundamental Researches (project \#96-01-00127, coordinator
Ya.T.~Sultanaev). This work also supported with grant of
Academy of Sciences of the Republic Bashkortostan
(coordinator of project N.M.~Asadullin).
\endthanks
\address
\noindent
Mathematical Department of Bashkir State University,
\newline Frunze street 32, 450074, Ufa, Russia.
\endaddress
\email
root\@bgua.bashkiria.su
\endemail
\abstract
Point transformations for the ordinary
differential equations of the form $y''=P(x,y)+3\,Q(x,y)\,y'+
3\,R(x,y)\,(y')^2+S(x,y)\,(y')^3$ are considered. Some
classical results are resumed. Solution for the equivalence
problem for the equations of general position is described.
\endabstract
\endtopmatter
\loadbold
\document
\head
1. Introduction.
\endhead
     Let's consider an ordinary differential equation of the
second order with the right hand side being cubic polynomial
in $y'$:
$$
y''=P(x,y)+3\,Q(x,y)\,y'+3\,R(x,y)\,(y')^2+S(x,y)\,(y')^3.
\tag1.1
$$
Class of the equations \thetag{1.1} is conserved under the
point transformations of the form
$$
\cases
\tilde x=\tilde x(x,y),\\
\tilde y=\tilde y(x,y).
\endcases
\tag1.2
$$
This means that after the change of variables \thetag{1.2}
in any one of such equations we shall obtain another equation
of the same form
$$
\tilde y''=\tilde P(\tilde x,\tilde y)+3\,\tilde Q(\tilde x,
\tilde y)\,\tilde y'+3\,\tilde R(\tilde x,\tilde y)\,(\tilde y')^2
+\tilde S(\tilde x,\tilde y)\,(\tilde y')^3.
\tag1.3
$$
If two particular equations \thetag{1.1} and \thetag{1.3}
are fixed, the question on the existence of the point
transformation \thetag{1.2} that transfer one of these
equations into another is known as {\it the problem of
equivalence}. In special case when the equation \thetag{1.3}
is trivial $\tilde y''=0$ this problem was solved in classical
papers by Tresse \cite{1} and Cartan \cite{2}.\par
     Another particular case for the equivalence problem is
connected with Painleve equations. These are six equations
of the form \thetag{1.1} with meromorphic coefficients
defined by the condition that their common solution
$y=y(x,c_1,c_2)$ considered as a function in $x$ has no
singularities except for the poles (see \cite{3} and \cite{4}).
Painleve equations became very popular with the advent
of the inverse scattering method, since they arise as
self-similar solutions for various equations, which are
integrable by this method. First two Painleve equations are
the following
$$
\align
&\tilde y''=\tilde y^2+\tilde x,
\tag1.4\\
&\tilde y''=\tilde y^3+\tilde x\,\tilde y+a,
\tag1.5
\endalign
$$
where $a=\const$. In \cite{5} the problem of equivalence for
the equations \thetag{1.4} and \thetag{1.5} was considered
for the point transformations of the special form:
$$
\cases
\tilde x=\tilde x(x),\\
\tilde y=\tilde y(x,y).
\endcases
\tag1.6
$$
Transformations \thetag{1.6} constitute a subset in the set
of general point transformations \thetag{1.2}. The generalization
of the result from \cite{5} for the case of arbitrary point
transformations \thetag{1.2} was obtained in \cite{6}.\par
     Main goal of our paper is to resume some classical
constructions from \cite{1} and \cite{2}, and apply them
to the solution of the equivalence problem for the equations
\thetag{1.1} being in general position. This is the most
broad class of equations of the form \thetag{1.1}, but
nevertheless many famous equations appear to be out of this
class. All six Painleve equations are not in general position,
therefore the solution of the equivalence for them requires
the separate consideration.\par
\head
2. Point transformations.
\endhead
     Let's suppose the point transformation \thetag{1.2} to be
regular. Denote by $T$ and $S$ direct and inverse matrices
of Jacoby for the transformation \thetag{1.2}
$$
\xalignat 2
&S=\Vmatrix
x_{\sssize 1.0} &x_{\sssize 0.1}\\
\vspace{1ex}
y_{\sssize 1.0} &y_{\sssize 0.1}
\endVmatrix,
&&T=\Vmatrix
\tilde x_{\sssize 1.0} &\tilde x_{\sssize 0.1}\\
\vspace{1ex}
\tilde y_{\sssize 1.0} &\tilde y_{\sssize 0.1}
\endVmatrix.
\tag2.1
\endxalignat
$$
By means of double indices in \thetag{2.1} and in what follows
we indicate partial derivatives. For the function $f(u,v)$
$f_{\sssize p.q}$ we denote the result of differentiation
$p$-times with respect to its first argument and $q$-times
with respect to the second argument.\par
     The formula for transforming the first order derivatives
by the point transforma\-tions \thetag{1.2} has the following form:
$$
y'=\frac{y_{\sssize 1.0}+y_{\sssize 0.1}\,\tilde y'}
{x_{\sssize 1.0}+x_{\sssize 0.1}\,\tilde y'}.
\tag2.2
$$
Analogous formula for the second order derivatives is written as
follows:
$$
\aligned
y''&=\frac{(x_{\sssize 1.0}+x_{\sssize 0.1}\,\tilde y')
(y_{\sssize 2.0}+2\,y_{\sssize 1.1}\,\tilde y'+y_{\sssize 0.2}
\,(\tilde y')^2+y_{\sssize 0.1}\,\tilde y'')}
{(x_{\sssize 1.0}+x_{\sssize 0.1}\,\tilde y')^3}-\\
\vspace{2ex}
&\qquad-\frac{(y_{\sssize 1.0}+y_{\sssize 0.1}
\,\tilde y')(x_{\sssize 2.0}+2\,x_{\sssize 1.1}\,\tilde y'+
x_{\sssize 0.2}\,(\tilde y')^2+x_{\sssize 0.1}\,\tilde y'')}
{(x_{\sssize 1.0}+x_{\sssize 0.1}\,\tilde y')^3}.
\endaligned
\tag2.3
$$
By substituting \thetag{2.2} and \thetag{2.3} into \thetag{1.1} we
define the transformation rule for the coefficients of the
equations \thetag{1.1} by the point transformation \thetag{1.2}.
In order to write this rule in a compact form let's construct a
three dimensional array with the following components determined
by the coefficients of the equation \thetag{1.1}
$$
\xalignat 2
&\theta_{111}=P,
&&\theta_{112}=\theta_{121}=\theta_{211}=Q,
\hskip -2em\\
\vspace{-1.7ex}
&&&\tag2.4\\
\vspace{-1.7ex}
&\theta_{122}=\theta_{212}=\theta_{221}=R,
&&\theta_{222}=S.
\hskip -2em
\endxalignat
$$
As we can see now from \thetag{2.4} the array $\theta_{ijk}$ is
symmetric in each pair of indices. Let's raise one of these
indices$$
\theta^k_{ij}=\sum^2_{r=1}d^{kr}\,\theta_{rij},
\tag2.5
$$
by means of contraction with the following skew-symmetric matrix
$d^{ij}$
$$
d_{ij}=d^{ij}=
\Vmatrix\format \r&\quad\l\\ 0 & 1\\-1 & 0\endVmatrix.
\tag2.6
$$
The transformation rule for the quantities \thetag{2.5} under
the point change of variables \thetag{1.2} can be written as
$$
\theta^k_{ij}=\sum^2_{m=1}\sum^2_{p=1}\sum^2_{q=1}
S^k_m\,T^p_i\,T^q_j\,\tilde\theta^m_{pq}
+\sum^2_{m=1}S^k_m\,\frac{\partial T^m_i}{\partial x^j}-
\frac{\tilde\sigma_i\,\delta^k_j+
\tilde\sigma_j\,\delta^k_i}{3},
\tag2.7
$$
where $x^1=x$, $x^2=y$, $\tilde x^1=\tilde x$, $\tilde x^2=\tilde y$
and where
$$
\xalignat 3
&\qquad\tilde\sigma_i=\frac{\partial\ln\det T}{\partial x^i},
&&\delta^k_i=
\cases
1&\text{for\ }i=k,\\
0&\text{for\ }i\neq k.
\endcases
\tag2.8
\endxalignat
$$
Because of the last summand with $\tilde\sigma_i$ and $\tilde
\sigma_j$ the formula \thetag{2.7} differs from the standard
rule of transformation for the components of connection
(see \cite{7}). But this shouldn't prevent us to construct
the four dimensional array of quantities which in Riemannian
geometry is known as a curvature tensor:
$$
\Omega^k_{rij}=
\frac{\partial\theta^k_{jr}}{\partial u^i}
-\frac{\partial\theta^k_{ir}}{\partial u^j}+
\sum^2_{q=1}\theta^k_{iq}\theta^q_{jr}-
\sum^2_{q=1}\theta^k_{jq}\theta^q_{ir}.
\tag2.9
$$
Under the local change of variables \thetag{1.2} the quantities
$\Omega^k_{rij}$ in \thetag{2.9} are transformed according to
the rule
$$
\Omega^k_{rij}=\sum^2_{m=1}\sum^2_{n=1}\sum^2_{p=1}\sum^2_{q=1}
S^k_m\,T^n_r\,T^p_i\,T^q_j\,\tilde\Omega^m_{npq}-
\frac{\tilde\sigma_{ir}\,\delta^k_j-
\tilde\sigma_{jr}\,\delta^k_i}{3},
\tag2.10
$$
which is different from the rule of transformation for the
components of tensor. The quantities $\tilde\sigma_{ij}$
in \thetag{2.10} are determined by $\tilde\sigma_i$
from \thetag{2.8} according to the formula
$$
\tilde\sigma_{ij}=\frac{\partial\tilde\sigma_j}
{\partial x^i}-\sum^2_{q=1}\theta^q_{ij}\,\tilde\sigma_q-
\frac{1}{3}\,\tilde\sigma_i\,\tilde\sigma_j.
\tag2.11
$$
Now let's contract the array $\Omega^k_{rij}$ by the pair
of indices $k$ and $i$
$$
\Omega_{rj}=\sum^2_{k=1}\Omega^k_{rkj}.
\tag2.12
$$
In Riemannian geometry the result of contraction \thetag{2.12}
is known as the tensor of Ricci. But here we obtain the
two-dimensional array $\Omega_{rj}$ which is not a tensor.
Under the point transformations \thetag{1.2} the quantities
$\Omega_{rj}$ are transformed as follows
$$
\Omega_{rj}=\sum^2_{n=1}\sum^2_{q=1}T^n_r\,T^q_j\,
\tilde\Omega_{nq}+\frac{1}{3}\,\tilde\sigma_{jr}.
\tag2.13
$$
Formula \thetag{2.13} is derived from \thetag{2.10}. Here we
deal with two dimensional manifold --- coordinate plane $(x,y)$.
Because of this two-dimensionality all components of the
array $\Omega^k_{rij}$ can be recovered from $\Omega_{rj}$:
$$
\Omega^k_{rij}=\Omega_{rj}\,\delta^k_i-
\Omega_{ri}\,\delta^k_j.
\tag2.14
$$\par
     Array of quantities $\Omega_{ij}$ is symmetric in $i$ and $j$.
Let's write down the values of its components in explicit form
$$
\aligned
&\Omega_{12}=\Omega_{21}=R_{\sssize 1.0}-
Q_{\sssize 0.1}+P\,S-Q\,R,\\
\vspace{1ex}
&\Omega_{11}=Q_{\sssize 1.0}-P_{\sssize 0.1}+
2\,P\,R-2\,Q^2,\\
\vspace{1ex}
&\Omega_{22}=S_{\sssize 1.0}-R_{\sssize 0.1}+
2\,S\,Q-2\,R^2.
\endaligned
\tag2.15
$$
Now we calculate derivatives $\nabla_i\Omega_{jk}$, using
$\theta^k_{ij}$ as components of connection
$$
\nabla_i\Omega_{jk}=\frac{\partial\Omega_{jk}}{\partial x^i}-
\sum^2_{r=1}\theta^r_{ij}\,\Omega_{rk}-\sum^2_{r=1}\theta^r_{ik}\,
\Omega_{jr}.
\tag2.16
$$
and by \thetag{2.16} we construct another three-dimensional
array $W_{ijk}$ skew-symmetric in first pair of indices
$$
W_{ijk}=\nabla_i\Omega_{jk}-\nabla_j\Omega_{ik}.
\tag2.17
$$
Quantities \thetag{2.16} don't form a tensor, the use of the sign
of covariant derivative $\nabla_i$ in \thetag{2.16} is quite formal.
However the quantities \thetag{2.17} do form a tensor. This
significant fact can be checked by direct calculations which are
based on \thetag{2.10}, \thetag{2.11} and \thetag{2.13}.\par
     Because of skew-symmetry in $i$ and $j$ the number of
nonzero components of tensor $W$ is $2$. The most simple way
to extract them is to contract $W_{ijk}$ with the matrix
$d^{ij}$ defined by the formula \thetag{2.6}
$$
\alpha_k=\frac{1}{2}\sum^2_{i=1}\sum^2_{j=1}W_{ijk}\,d^{ij}.
\tag2.18
$$
Quantities $\alpha_1$ and $\alpha_2$ can be calculated
directly from the coefficients of the equation \thetag{1.1}.
Let's write down the appropriate formulas
$$
\aligned
&\aligned
 A=\alpha_1=P_{\sssize 0.2}&-2\,Q_{\sssize 1.1}+R_{\sssize 2.0}+
 2\,P\,S_{\sssize 1.0}+S\,P_{\sssize 1.0}-\\
 \vspace{0.5ex}
 &-3\,P\,R_{\sssize 0.1}-3\,R\,P_{\sssize 0.1}-
 3\,Q\,R_{\sssize 1.0}+6\,Q\,Q_{\sssize 0.1},
 \endaligned\\
 \vspace{1ex}
&\aligned
 B=\alpha_2=S_{\sssize 2.0}&-2\,R_{\sssize 1.1}+Q_{\sssize 0.2}-
 2\,S\,P_{\sssize 0.1}-P\,S_{\sssize 0.1}+\\
 \vspace{0.5ex}
 &+3\,S\,Q_{\sssize 1.0}+3\,Q\,S_{\sssize 1.0}+
 3\,R\,Q_{\sssize 0.1}-6\,R\,R_{\sssize 1.0}.
 \endaligned
\endaligned
\tag2.19
$$
Components $d^{ij}$ do not form a tensor, therefore the result of
contraction \thetag{2.18} is not a tensor too. Arrays $\alpha_k$
and $d^{ij}$ belong to special class geometrical objects which
are known as {\it pseudotensors}.
\definition{Definition 2.1} Pseudotensorial field of the type
$(r,s)$ and weight $m$ is an array of quantities $F^{i_1\ldots\,
i_r}_{j_1\ldots\,j_s}$ which under the change of variables
\thetag{1.2} transforms as follows
$$
F^{i_1\ldots\,i_r}_{j_1\ldots\,j_s}=
(\det T)^m\sum\Sb p_1\ldots p_r\\ q_1\ldots q_s\endSb
S^{i_1}_{p_1}\ldots\,S^{i_r}_{p_r}\,\,
T^{q_1}_{j_1}\ldots\,T^{q_s}_{j_s}\,\,
\tilde F^{p_1\ldots\,p_r}_{q_1\ldots\,q_s}.
\tag2.20
$$
\enddefinition
     Traditional tensorial fields can be treated as pseudotensorial
fields of the weight $m=0$ in \thetag{2.20}. Quantities $d^{ij}$ in
\thetag{2.6} and quantities $\alpha_k$ in \thetag{2.18} form the
components of pseudotensorial fields of the weight $m=1$. As an
important consequence of pseudotensorial character of the quantities
$A$ and $B$ in \thetag{2.19} we get the following lemma.
\proclaim{Lemma 2.1} If both parameters $A$ and $B$ are zero $A=B=0$
for the initial equation \thetag{1.1}, then they both are zero
$\tilde A=0$ and $\tilde B=0$ for the transformed equation
\thetag{1.3}.
\endproclaim
     Now let's calculate $\nabla_i\alpha_k$ using $\theta^k_{ij}$
from \thetag{2.5} as the components of connection. Then determine
the quantities $\beta_i$ according to the formula
$$
\beta_i=3\,\sum^2_{k=1}\sum^2_{r=1}\nabla_i\alpha_k\,d^{kr}\,
\alpha_r+\sum^2_{k=1}\sum^2_{r=1}\nabla_r\alpha_k
\,d^{kr}\,\alpha_i.
\tag2.21
$$
Quantities \thetag{2.21} appear to be the components of
pseudocovectorial field of the weight $3$. Raising indices
in $\alpha_i$ and $\beta_i$ we get two pseudovectorial
fields with the weights $2$ and $4$ respectively
$$
\xalignat 2
&\alpha^i=\sum^2_{k=1}d^{ik}\,\alpha_k,
&
&\beta^i=\sum^2_{k=1}d^{ik}\,\beta_k,
\tag2.22
\endxalignat
$$
We can calculate $\beta^1$ and $\beta^2$ directly from the coefficients
of the equation \thetag{1.1}:
$$
\aligned
\beta^1=G&=-B\,B_{\sssize 1.0}-3\,A\,B_{\sssize 0.1}+4\,B\,
A_{\sssize 0.1}+3\,S\,A^2-6\,R\,B\,A+3\,Q\,B^2,\\
\vspace{1ex}
\beta^2=H&=-A\,A_{\sssize 0.1}-3\,B\,A_{\sssize 1.0}+4\,A\,
B_{\sssize 1.0}-3\,P\,B^2+6\,Q\,A\,B-3\,R\,A^2.
\endaligned
\tag2.23
$$
Now let's define the quantity $F$ by means of contraction of
the fields \thetag{2.18} and \thetag{2.22}
$$
3\,F^5=\sum^2_{i=1}\alpha_i\,\beta^i=-\sum^2_{i=1}\beta_i\,\alpha^i=
A\,G+B\,H.
\tag2.24
$$
We can also write down the explicit formula for $F$
$$
\aligned
F^5=A\,B\,A_{\sssize 0.1}&+B\,A\,B_{\sssize 1.0}-
A^2\,B_{\sssize 0.1}-B^2\,A_{\sssize 1.0}-\\
&-P\,B^3+3\,Q\,A\,B^2-3\,R\,A^2\,B+S\,A^3.
\endaligned
\tag2.25
$$
The quantity $F$ given by the formula \thetag{2.25} is a pseudoscalar
field of the weight $1$. As an immediate consequence of this fact we
obtain the following lemma.
\proclaim{Lemma 2.2} Vanishing of the parameter $F=0$ for initial
equation \thetag{1.1} is equivalent to vanishing $\tilde F=0$ for
the transformed equation \thetag{1.3}.
\endproclaim
    Lemma~2.2 separate two quite different cases in the study of
the equations \thetag{1.1}: $F=0$ and $F\neq 0$. The case $F\neq 0$
appears to be more structured from the geometrical point of view
--- it is the case of general position.
\head
3. Case of general position.
\endhead
    Let $F\neq 0$. This means that pseudovectorial fields $\alpha$
and $\beta$ are not collinear. We can form two vectorial fields
$\bold X$ and $\bold Y$ from them
$$
\xalignat 2
&X^i=\frac{\alpha^i}{F^2},
&&Y^i=\frac{\beta^i}{F^4}.
\tag3.1
\endxalignat
$$
Nonzero pseudoscalar field $F$ lets us to define the quantities
$$
\varphi_i=-\frac{\partial\ln F}{\partial x^i},
\tag3.2
$$
which under the point change of variables \thetag{1.2} should
transform as follows
$$
\varphi_i=\sum^2_{j=1}T^j_i\,\tilde\varphi_j-\sigma_i.
\tag3.3
$$
Compare this rule of transformation for the quantities \thetag{3.2}
with the rule of trans\-formation for $\theta^k_{ij}$. Such comparison
of \thetag{3.3} with \thetag{2.10} enables us to modify the quantities
$\theta^k_{ij}$ converting them into the components of an affine
connection
$$
\varGamma^k_{ij}=\theta^k_{ij}-\frac{\varphi_i\,\delta^k_j+
\varphi_j\,\delta^k_i}{3}
\tag3.4
$$
Noncollinear vector fields $\bold X$ and $\bold Y$ form the
moving frame in the coordinate plane $(x,y)$. Consider the
components of the connection \thetag{3.4}, related to this
frame. They are defined as coefficients $\Gamma^1_{11}$ in
the following expansions:
$$
\xalignat 2
&\nabla_{\bold X}\bold X=\Gamma^1_{11}\,\bold X+
\Gamma^2_{11}\,\bold Y,
&
&\nabla_{\bold X}\bold Y=\Gamma^1_{12}\,\bold X+
\Gamma^2_{12}\,\bold Y,\\
\vspace{-1.7ex}
&&&\tag3.5\\
\vspace{-1.7ex}
&\nabla_{\bold Y}\bold X=\Gamma^1_{21}\,\bold X+
\Gamma^2_{21}\,\bold Y,
&
&\nabla_{\bold Y}\bold Y=\Gamma^1_{22}\,\bold X+
\Gamma^2_{22}\,\bold Y.
\endxalignat
$$
In contrast to the quantities $\varGamma^k_{ij}$ in \thetag{3.4}
the coefficients $\Gamma^k_{ij}$ in \thetag{3.5} do not change
under the point transformations \thetag{1.2}. They are scalar
fields, i.e. they are {\it scalar invariants} of the equation
\thetag{1.1}. Let's numerate them as follows: $I_1=\Gamma^1_{11}$,
$I_2=\Gamma^2_{11}$, $I_3=\Gamma^1_{12}$, \dots, $I_8=\Gamma^2_{22}$.
List of scalar invariants of the equation \thetag{1.1} can be
continued. We can differentiate these invariants along the vector
fields $\bold X$ and $\bold Y$ obtaining as a result more and more
new invariants: $I_9=\bold XI_1$, \dots, $I_{16}=\bold XI_8$,
$I_{17}=\bold YI_1$, \dots, $I_{24}=\bold YI_8$. Repeating this
procedure gives us $16$ invariants in each step. The general
structure of invariants distinguishes three different cases
\roster
\item in the infinite sequence of invariants $I_1$, $I_2$, $I_3$,
      \dots there is a pair of functionally independent ones;
\item all invariant in the sequence $I_1$, $I_2$, $I_3$, \dots
      are functionally dependent, but not all are identically
      constant;
\item all invariants in the sequence $I_1$, $I_2$, $I_3$, \dots
      are constants.
\endroster
\par
    {\bf Case 1.} Consider the set of pairs of invariants arranged
in a lexicographic ordering. Then choose the first pair of
functionally independent invariants in this ordering: $(I_p,I_q)$.
We can use $I_p(x,y)$ and $I_q(x,y)$ in order to define point
transformation
$$
\cases
\tilde x=I_p(x,y),\\
\tilde y=I_q(x,y).
\endcases
\tag3.6
$$
Variables $\tilde x$ and $\tilde y$ defined by invariants $I_p$ and
$I_q$ are natural to be taken for the {\it canonical variables}
of the equation \thetag{1.1}. The equation \thetag{1.3} obtained
as a result of transformation \thetag{3.6} is natural to call
the {\it canonical form} of the equation \thetag{1.1}. Solution
of the equivalence problem in this case is given by the
following obvious theorem.
\proclaim{Theorem 3.1} Two equations of the form \thetag{1.1} with
nonzero parameters $F$ and with functionally independent invariants
are equivalent if and only if they have the same canonical form.
\endproclaim
     Before considering the second case let's write down explicit
formulas for the first eight invariants $I_1$, $I_2$, \dots, $I_8$.
For $I_1$ and $I_3$ we have
$$
\align
&\hskip -1em\aligned
 I_1&=\frac{B\,(G\,A_{\sssize 1.0}+H\,B_{\sssize 1.0})}{3\,F^7}
 -\frac{A\,(G\,A_{\sssize 0.1}+H\,B_{\sssize 0.1})}{3\,F^7}+
 \frac{4\,(A\,F_{\sssize 0.1}-B\,F_{\sssize 1.0})}{3\,F^3}+\\
 \vspace{1ex}
 &+\frac{G\,B^2\,P}{3\,F^7}+\frac{(H\,B^2-2\,G\,B\,A)\,Q}{3\,F^7}+
 \frac{(G\,A^2-2\,H\,B\,A)\,R}{3\,F^7}+\frac{H\,A^2\,S}{3\,F^7}
 \endaligned\hskip -4em
\tag3.7\\
\vspace{3ex}
&\hskip -1em\aligned
 I_3&=\frac{B\,(H\,G_{\sssize 1.0}-G\,H_{\sssize 1.0})}{3\,F^9}
 -\frac{A\,(H\,G_{\sssize 0.1}-G\,H_{\sssize 0.1})}{3\,F^9}+
 \frac{H\,F_{\sssize 0.1}+G\,F_{\sssize 1.0}}{3\,F^5}+\\
 \vspace{1ex}
 &+\frac{B\,G^2\,P}{3\,F^9}-\frac{(A\,G^2-2\,H\,B\,G)\,Q}{3\,F^9}+
 \frac{(B\,H^2-2\,H\,A\,G)\,R}{3\,F^9}-\frac{A\,H^2\,S}{3\,F^9}
 \endaligned\hskip -4em
\tag3.8
\endalign
$$
Formula for $I_7$ has much more simple form than formulas \thetag{3.7}
and \thetag{3.8}
$$
\aligned
 I_7=&\frac{G\,H\,G_{\sssize 1.0}-
 G^2\,H_{\sssize 1.0}+H^2\,G_{\sssize 0.1}-
 H\,G\,H_{\sssize 0.1}}{3\,F^{11}}+\\
 \vspace{1ex}
 &\hskip 5em+\frac{G^3\,P+3\,G^2\,H\,\,Q+3\,G\,H^2\,R+H^3\,S}
 {3\,F^{11}}
 \endaligned
\tag3.9
$$
Invariant $I_2$ require no calculations --- it is simply an
identical constant
$$
I_2=\frac{1}{3}.
\tag3.10
$$
In order to calculate rest four invariants let's define the following
two quantities which are also scalar invariants
$$
\align
K&=I_6-I_4=\Gamma^2_{21}-\Gamma^2_{12}=
\frac{1}{F}\frac{\partial}{\partial x}\fracwithdelims(){B}{F}-
\frac{1}{F}\frac{\partial}{\partial y}\fracwithdelims(){A}{F},
\tag3.11\\
\vspace{1ex}
L&=I_3-I_5=\Gamma^1_{12}-\Gamma^1_{21}=
\frac{1}{F}\frac{\partial}{\partial x}\fracwithdelims(){G}{F^3}+
\frac{1}{F}\frac{\partial}{\partial y}\fracwithdelims(){H}{F^3}.
\tag3.12
\endalign
$$
Then invariants $I_4$, $I_5$, $I_6$ and $I_8$ are defined by
the following relationships
$$
\xalignat 2
&I_4=\Gamma^2_{12}=-I_1,
&
&I_5=\Gamma^1_{21}=I_3-L,
\tag3.13\\
&I_6=\Gamma^2_{21}=-I_1+K,
&
&I_8=\Gamma^2_{22}=-I_5.
\tag3.14
\endxalignat
$$
The quantities $K$ and $L$ in \thetag{3.11} and \thetag{3.12}
have clear geometrical interpretation due to the identity
$[\bold X,\,\bold Y]=\nabla_{\bold X}\bold Y-\nabla_{\bold Y}
\bold X$ (see for instance \cite{8}):
$$
[\bold X,\,\bold Y]=L\,\bold X-K\,\bold Y.
\tag3.15
$$
Choose the functions $u$ and $v$ so that commutator of the
following vector fields
$$
\xalignat 2
&\tilde\bold X=\frac{\bold X}{u},
&
&\tilde\bold Y=\frac{\bold Y}{v},
\tag3.16
\endxalignat
$$
be equal to zero. Because of \thetag{3.15} this condition
gives two differential equations for the functions $u$ and
$v$ in \thetag{3.16}
$$
\xalignat 2
&\bold Yu=-L\,u,
&
&\bold Xv=-K\,v.
\tag3.17
\endxalignat
$$
Written in coordinates the equations \thetag{3.17} appears
to be linear differential equations of the first order.
Such equations are solved by means of the method of
characteristics (see \cite{9}). Thus we can choose pair
of nonzero functions $u$ and $v$ that satisfy the condition
$[\tilde\bold X,\,\tilde\bold Y]=0$ for the vector fields
\thetag{3.16}.\par
   Any pair of commutating vector fields on a plane defines
some curvilinear system of coordinates $(\tilde x,\tilde y)$
for which $\tilde\bold X$ and $\tilde\bold Y$ form the
coordinate frame. Let's perform the point transformation
to the coordinates $\tilde x$ and $\tilde y$ defined by
$\tilde\bold X$ and $\tilde\bold Y$. Here
$$
\xalignat 2
&\tilde\alpha^1=\tilde B=u\,\tilde F^2,
&&\tilde\alpha^2=-\tilde A=0,\\
\vspace{-1.2ex}
&&&\tag3.18\\
\vspace{-1.2ex}
&\tilde\beta^1=\tilde G=0,
&&\tilde\beta^2=\tilde H=v\,\tilde F^4,
\endxalignat
$$
Substituting $\tilde A=0$ and $\tilde G=0$ from \thetag{3.18} into
\thetag{2.23} we get the following equations:
$$
\xalignat 2
&3\,\tilde Q\,\tilde B^2-\tilde B\,\tilde B_{\sssize 1.0}=0,
&
&\tilde H=-3\,\tilde P\,\tilde B^2.
\tag3.19
\endxalignat
$$
Taking into account \thetag{2.24} and the relationships
$\tilde B=u\,\tilde F^2$ and $\tilde H=v\,\tilde F^4$
we can express $\tilde P$, $\tilde Q$ and the functions
$u$ and $v$ via $\tilde B$ and $\tilde F$
$$
\xalignat 4
&\ u=\frac{\tilde B}{\tilde F^2},
&&v=\frac{3\,\tilde F}{\tilde B},
&&\tilde P=-\frac{\tilde F^5}{\tilde B^3},
&&\tilde Q=\frac{\tilde B_{\sssize 1.0}}{3\,\tilde B}.
\hskip -2em
\tag3.20
\endxalignat
$$
Take nonzero parameters $\tilde B\neq 0$ and $\tilde F\neq 0$
for the basic ones in order to express invariants $I_1$, $I_3$,
$I_7$, $K$ and $L$ in terms of them
$$
\xalignat 3
&\qquad I_1=\frac{4}{3}\,\frac{\tilde F\,\tilde B_{\sssize 1.0}-
\tilde B\,\tilde F_{\sssize 1.0}}{\tilde F^3},
&&I_3=\frac{\tilde F_{\sssize 0.1}+3\,\tilde F\,\tilde R}{\tilde B},
&&I_7=\frac{9\,\tilde F^4\,\tilde S}{\tilde B^3}.
\tag3.21
\endxalignat
$$
For the invariants $K$ and $L$ that determine the commutator of
$\bold X$ and $\bold Y$ we get the following expressions
$$
\xalignat 2
&K=\frac{\tilde F\,\tilde B_{\sssize 1.0}-
\tilde B\,\tilde F_{\sssize 1.0}}{\tilde F^3},
&&L=\frac{6\,\tilde B\tilde F_{\sssize 0.1}-3\,\tilde F\,
\tilde B_{\sssize 0.1}}{\tilde B^2}.
\tag3.22
\endxalignat
$$
Invariants $I_4$, $I_5$, $I_6$, $_8$ are determined by the
relationships \thetag{3.13} and \thetag{3.14}. In addition to
these relationships from comparison of \thetag{3.21} and
\thetag{3.22} we derive
$$
I_1=\frac{4}{3}\,K.
\tag3.23
$$
Note that the relationship \thetag{3.23} can be derived directly
from \thetag{3.7} and \thetag{3.11}. However this require more
complicated calculations.\par
    {\bf Case $2$.} In this case all invariants $I_k$ are
functionally dependent but not all of them are constants. Find
first nonconstant invariant $I=I_p$ among them. Then all invariants
$I_k$ can be expressed as some functions of $I$, i.e. $I_k=I_k(I)$.
The same is true for invariants $K$ and $L$ in \thetag{3.17}.
We shall seek the solutions for the equations \thetag{3.17} in
form of $u=u(I)$ and $v=v(I)$
$$
\xalignat 2
&u'(I)\,\bold YI=-L(I)\,u(I),
&
&v'(I)\,\bold XI=-K(I)\,v(I).
\tag3.24
\endxalignat
$$
Both derivatives of $I$ along the vector fields $\bold X$ and
$\bold Y$ are also the invariants in the sequence $I_1$, $I_2$,
$I_3$, \dots, therefore $\bold XI=\xi(I)$ and $\bold YI=\zeta(I)$.
Substituting this into \thetag{3.7} we bring the equations
\thetag{3.24} to the form of ordinary differential equations
for the functions $u(I)$ and $v(I)$
$$
\xalignat 2
&u'\,\zeta(I)=-L(I)\,u,
&
&v'\,\xi(I)=-K(I)\,v.
\tag3.25
\endxalignat
$$
For $\xi(I)\neq 0$ and $\zeta(I)\neq 0$ the equations \thetag{3.25}
are obviously solvable. Their solutions can be chosen nonzero
$u(I)\neq 0$ and $v(I)\neq 0$. Because of $I\neq\const$ and because
of linear independence of the vectors $\bold X$ and $\bold Y$ the
functions $\bold XI=\xi(I)$ and $\bold YI=\zeta(I)$ cannot vanish
simultaneously. Let's differentiate the invariant $I$ along the
commutator $[\bold X,\,\bold Y]$ of the vector fields $\bold X$
and $\bold Y$
$$
[\bold X,\,\bold Y]I=\bold X(\bold YI)-
\bold Y(\bold XI)=
\zeta'(I)\,\bold XI-\xi'(I)\,\bold YI=
\zeta'(I)\,\xi(I)-\xi'(I)\,\zeta(I).
$$
On the other hand due to \thetag{3.15} for the same expression
$[\bold X,\,\bold Y]I$ we get
$$
[\bold X,\,\bold Y]I=L(I)\,\bold XI-K\,\bold YI=
L(I)\,\xi(I)-K(I)\,\zeta(I).
$$
Combining these two relationships for $[\bold X,\,\bold Y]I$ we
derive the following equality
$$
\zeta'(I)\,\xi(I)-\xi'(I)\,\zeta(I)=
L(I)\,\xi(I)-K(I)\,\zeta(I).
\tag3.26
$$
On the base of \thetag{3.26} it's easy to find that $\xi(I)=0$
leads to $K(I)=0$, and $\zeta(I)=0$ leads to $L(I)=0$. Therefore
if $\xi(I)=0$, we can take $v(I)=1$, and if $\zeta(I)=0$, we
can take $u(I)=1$. This let's us satisfy the equations \thetag{3.24}
and \thetag{3.25} in any case.\par
     The pair of commutating vector fields \thetag{3.16} defined
by the choice of functions $u(I)$ and $v(I)$ determines the
choice of curvilinear coordinates $\tilde x$ and $\tilde y$ on
the plane such that $\tilde A=0$ and $\tilde G=0$. For the
parameters $\tilde F$ and $\tilde B$ from \thetag{3.20} we get
$$
\xalignat 2
&\tilde F=3\,u^{-1}\,v^{-1},
&&\tilde B=9\,u^{-1}\,v^{-2}.
\tag3.27
\endxalignat
$$
From \thetag{3.27} we see that $\tilde F=\tilde F(I)$ and $\tilde B=
\tilde B(I)$. For the parameters \thetag{3.2} this gives
$$
\xalignat 2
&\varphi_1=-\frac{F'(I)\,\xi(I)}{u(I)\,F(I)},
&&\varphi_2=-\frac{F'(I)\,\zeta(I)}{v(I)\,F(I)}.
\tag3.28
\endxalignat
$$
Hence parameters $\varphi_i$ are also functions in $I$. In the
coordinates $\tilde x$ and $\tilde y$ the components of connection
\thetag{3.4} can be expressed via the coefficients of the
expansions \thetag{3.5} which are the scalar invariants
$$
\xalignat 2
&\quad\tilde\varGamma^1_{11}=u^{-1}\,\Gamma^1_{11}-
\xi\,u^{-2}\,u',
&&\tilde\varGamma^2_{11}=v\,u^{-2}\,\Gamma^2_{11},\\
\vspace{1ex}
&\quad\tilde\varGamma^1_{12}=\tilde\varGamma^1_{21}=
v^{-1}\,\Gamma^1_{12},
&&\tilde\varGamma^2_{12}=\tilde\varGamma^2_{21}=
u^{-1}\,\Gamma^2_{21},
\tag3.29\\
\vspace{1ex}
&\quad\tilde\varGamma^1_{22}=u\,v^{-2}\,\Gamma^1_{22},
&&\tilde\varGamma^2_{22}=v^{-1}\,\Gamma^2_{22}-
\zeta\,v^{-2}\,v'.\hskip -2em
\endxalignat
$$
In order to transfer from $\tilde\varGamma^k_{ij}$ to $\tilde
\theta_{kij}$ we should perform an operation inverse to the
raising of index in \thetag{2.5}. This is done by means of
the matrix $-d_{kp}$
$$
\tilde\theta_{kij}=-\sum^2_{p=1}d_{kp}\,\tilde\varGamma^p_{ij}
-\frac{\tilde\varphi_i\,d_{kj}+\tilde\varphi_j\,d_{ki}}{3}.
\tag3.30
$$
But the quantities $\tilde\theta_{kij}$ coincide with the
coefficients of the equation \thetag{1.3} in coordinates
$\tilde x$ and $\tilde y$. Therefore from \thetag{3.28},
\thetag{3.29} and \thetag{3.30} we obtain
$$
\xalignat 4
&\ \tilde P=\tilde P(I), &&\tilde Q=\tilde Q(I),
&&\tilde R=\tilde R(I), &&\tilde S=\tilde S(I).
\hskip -2em
\tag3.31
\endxalignat
$$
Conclusion: coefficients of the transformed equation \thetag{1.3}
and its parameters $\tilde B$ are $\tilde F$ the functions in $I$.
Invariant $I$ in turn is a function in $\tilde x$ and $\tilde y$.
Let's study the dependence of $I$ on $\tilde x$ and $\tilde y$.
In order to do it let's consider the derivatives
$$
\xalignat 2
&\ I_{\sssize 1.0}=\frac{\bold XI}{u}=\frac{\xi(I)}{u(I)}=h(I),
&&I_{\sssize 0.1}=\frac{\bold YI}{v}=\frac{\zeta(I)}{v(I)}=
k(I).
\tag3.32
\endxalignat
$$
and calculate the second order derivative $I_{\sssize 1.1}$
from \thetag{3.32}. This can be done in two ways, therefore
we get the relationship
$$
h'(I)\,k(I)-k'(I)\,h(I)=0,
\tag3.33
$$
which should be considered as the compatibility condition for
the equations \thetag{3.32}. The relationship \thetag{3.33}
can be integrated. This gives the linear dependence of the
functions $h(I)$ and $k(I)$, i.e. there are two constants $C_1$
and $C_2$ such that
$$
C_2\,h(I)-C_1\,k(I)=0.
\tag3.34
$$
The relationship \thetag{3.34} is the equation for the function $I$
$$
C_2\,\frac{\partial I}{\partial\tilde x}-
C_1\,\frac{\partial I}{\partial\tilde y}=0.
\tag3.35
$$
The equation \thetag{3.35} is easily integrable by means of the
method of characteristics. Denote $\tau=C_1\,\tilde x+C_2\,\tilde y$.
Then the general solution of the differential equation \thetag{3.35}
is given by an arbitrary function of one variable $\tau$
$$
I=I(\tau)=I(C_1\,\tilde x+C_2\,\tilde y).
\tag3.36
$$
Note that the functions $u(I)$ and $v(I)$ are determined by the
differential equations \thetag{3.25} only up to a constant factor.
This let's us make the constants $C_1$ and $C_2$ in \thetag{3.36}
equal to unity if their initial values are not zero. Therefore
the general form of dependence of $\tau$ on $\tilde x$ and $\tilde y$
can be reduced to the following special cases
$$
\xalignat 3
&\tau=\tilde x+\tilde y,
&&\tau=\tilde x,
&&\tau=\tilde y.
\tag3.38
\endxalignat
$$
In any of these three cases defined by \thetag{3.38}, we can
start by choosing two arbitrary nonzero functions $\tilde F(\tau)$
and $\tilde B(\tau)$. Then define the coefficients $\tilde P(\tau)$
and $\tilde Q(\tau)$ for the equation \thetag{1.3} by means of
formulas \thetag{3.20}. And finally define the rest two
coefficients $\tilde R(\tau)$ and $\tilde S(\tau)$ for the
equation \thetag{1.3} by solving the system of ordinary
differential equations derived from \thetag{2.19}. For the case
$\tau=\tilde x+\tilde y$ this system is as follows
$$
\aligned
&P''-2\,Q''+R''+2\,P\,S'+(S-3\,R)\,P'-3\,(P+Q)\,R'+6\,Q\,Q'=0,\\
\vspace{1ex}
&S''-2\,R''+Q''-2\,S\,P'-(P-3\,Q)\,S'+3\,(S+R)\,Q'-6\,R\,R'=B.
\endaligned
\tag3.39
$$
When $\tau=\tilde x$ the system of equations \thetag{3.39} should be
replaced by the following one
$$
\aligned
&R''+2\,P\,S'+S\,P'-3\,Q\,R'=0,\\
\vspace{1ex}
&S''+3\,Q\,S'+3\,S\,Q'-6\,R\,R'=B.
\endaligned
\tag3.40
$$
In the last third case $\tau=\tilde y$ from \thetag{3.20} we have
$\tilde Q=0$. Therefore the system of equations for $\tilde R$
and $\tilde S$ here is even simpler than \thetag{3.40}
$$
\aligned
&P''+S\,P'-3\,S\,R'-3\,R\,P'=0,\\
\vspace{1ex}
&-2\,S\,P'-P\,S'=B.
\endaligned
\tag3.41
$$
The above procedure of choosing the coefficients of the equation
\thetag{1.3} based on the equations \thetag{3.39}, \thetag{3.40}
and \thetag{3.41} gives the complete description of the canonical
form of the equations \thetag{1.1} for the case of functionally
dependent invariants.\par
     {\bf Case $3$.} Remember that in this case all invariant in
the sequence $I_1$, $I_2$, $I_3$, \dots are identical constants.
Indeed we can check that first eight of them are constants. Then
all other invariants $I_{9}$, $I_{10}$, \dots will be zero. In
this case we can also construct the commutating vector fields
\thetag{3.16} and choose curvilinear coordinates $\tilde x$ and
$\tilde y$ defined by them. Functions $u$ and $v$ are the
solutions of the equations \thetag{3.17} where now $K=\const$
and $L=\const$. These equations admit some arbitrariness in the
choice of their solutions. We shall use this arbitrariness in
order to make the equation \thetag{1.3} as simple as possible.
\par
     Let's show that invariants $K$ and $L$ cannot vanish
simultaneously. If $K=L=0$, then we can choose $u=v=1$ and for
the parameters $\tilde F$, $\tilde B$, $\tilde P$ and $\tilde Q$
from \thetag{3.20} we derive
$$
\xalignat 4
&\qquad\tilde F=3,
&&\tilde B=9,
&&\tilde P=-\frac{1}{3},
&&\tilde Q=0.
\tag3.42
\endxalignat
$$
Substituting \thetag{3.42} in \thetag{3.21} we can express
$\tilde R$ and $\tilde S$ through invariants $I_3$ and $I_7$:
$$
\xalignat 2
&\qquad\tilde R=I_3,
&&\tilde S=I_7.
\tag3.43
\endxalignat
$$
Because of \thetag{3.42} and \thetag{3.43} and because of
constancy of invariants $I_1$, \dots, $I_8$ all coefficients
in \thetag{1.3} should be the constants. Substituting them
in \thetag{2.19} we get $\tilde B=0$. This contradict to the
equality $\tilde B=9$ from \thetag{3.42}.\par
     First case is $K=0$ and $L\neq 0$. Here we choose $v=1$.
The choice of $u$ we implement in two steps. First we choose
an arbitrary solution for the equation $\bold Yu=-L\,u$
from \thetag{3.17}. Denote this preliminary choice by $\hat u$.
It defines the curvilinear coordinates in which the differential
equation $\bold Yu=-L\,u$ for $u$ has the form:
$$
\frac{\partial u}{\partial\hat y}=-L\,u.
\tag3.44
$$
Being the solution of the equation \thetag{3.44} the function
$\hat u$ has the form $\hat u(\hat x,\hat y)=\hat u(\hat x)\,
e^{-L\hat y}$. Now we take one more solution for the equation
\thetag{3.44} given by the formula $u(\hat x,\hat y)=e^{-L\hat y}$.
This ultimate choice of functions $u=e^{-L\hat y}$ and $v=1$
determines new coordinates $\tilde x=f(\hat x)$ and $\tilde y=
\hat y$ in which the function $u$ has the form $u(\tilde x,
\tilde y)=e^{-L\tilde y}$ from the very beginning. In these
canonical coordinates from \thetag{3.20} we derive
$$
\xalignat 4
&\qquad\tilde F=3\,e^{L\tilde y},
&&\tilde B=9\,e^{L\tilde y},
&&\tilde P=-\frac{1}{3}\,e^{2L\tilde y},
&&\tilde Q=0.
\tag3.45
\endxalignat
$$
Then on the base of \thetag{3.21} we express $\tilde R$ and $\tilde S$
through the invariants $I_3$ and $I_7$
$$
\xalignat 2
&\qquad\tilde R=I_3-\frac{L}{3},
&&\tilde S=I_7\,e^{-L\tilde y}.
\tag3.46
\endxalignat
$$
By substituting \thetag{3.45} and \thetag{3.46} into \thetag{2.19}
we find $I_3=L$ and $I_7=9/L$. Therefore the equation \thetag{1.3}
is written as
$$
\tilde y''=-\frac{1}{3}\,e^{2L\tilde y}+2\,L\,(\tilde y')^2+
\frac{9}{L}\,e^{-L\tilde y}\,(\tilde y')^3.
\tag3.47
$$
The equation \thetag{3.47} is a canonical form for the equation
\thetag{1.1} with identically constant invariants when $F\neq 0$,
$K=0$ and $L\neq 0$.\par
     Now we consider another case $L=0$ and $K\neq 0$. Take $u=1$
and choose the function $v$ satisfying the equation $\bold Xv=
-K\,v$ from \thetag{3.17} such that in canonical coordinates
$\tilde x$ and $\tilde y$ it has the form $v=e^{-K\tilde x}$.
We omit the details of such choice since it's quite similar
to the choice of $u$ in previous case. From \thetag{3.20} we
determine the parameters $\tilde F$, $\tilde B$, $\tilde P$
and $\tilde Q$ in canonical coordinates
$$
\xalignat 4
&\qquad\tilde F=3\,e^{K\tilde x},
&&\tilde B=9\,e^{2K\tilde x},
&&\tilde P=-\frac{1}{3}\,e^{-K\tilde x},
&&\tilde Q=\frac{2}{3}\,K.
\tag3.48
\endxalignat
$$
Coefficients $\tilde R$ and $\tilde S$ are expressed through
$I_3$ and $I_7$ by means of \thetag{3.21}
$$
\xalignat 2
&\qquad\tilde R=I_3\,e^{K\tilde x},
&&\tilde S=I_7\,e^{2K\tilde x}.
\tag3.49
\endxalignat
$$
By substituting \thetag{3.48} and \thetag{3.49} into the
relationships \thetag{2.19} and taking into account $\tilde A=0$
we get
$$
\aligned
&I_7+K\,I_3=0,\\
\vspace{1ex}
&-8\,I_7*K^2+6\,K*I_3^2+9=0.
\endaligned
\tag3.50
$$
Canonical form of the equation \thetag{1.1} in this case is
as follows
$$
\tilde y''=-\frac{1}{3}\,e^{-K\tilde x}+2\,K\,\tilde y'+
3\,I_3\,e^{K\tilde x}\,(\tilde y')^2+I_7\,e^{2K\tilde x}\,
(\tilde y')^3.
\tag3.51
$$
Here parameters $I_3$ and $I_7$ are defined by their parameter $K$
from the equations \thetag{3.50}.\par
     The rest case is $L\neq 0$ and $K\neq 0$. In this case we
should the functions $u$ and $v$ simultaneously. First we choose
two arbitrary functions satisfying the equations \thetag{3.17}.
They define curvilinear coordinates $\hat x$ and $\hat y$ in
which the equations \thetag{3.17} have the following form
$$
\xalignat 2
&v\,\frac{\partial u}{\partial\hat y}=-L\,u,
&&u\,\frac{\partial v}{\partial\hat x}=-K\,v.
\tag3.52
\endxalignat
$$
General solution for the system of equations \thetag{3.52}
is determined by two arbitrary functions in one variable
$\hat p(\hat x)$ and $\hat q(\hat y)$
$$
\xalignat 2
&\hat u=-K\,\frac{\hat p(\hat x)+\hat q(\hat y)}{\hat p'(\hat x)},
&&\hat v=-L\,\frac{\hat p(\hat x)+\hat q(\hat y)}{\hat q'(\hat y)}.
\tag3.53
\endxalignat
$$
The following special point transformation $\hat x=f(\tilde x)$ and
$\hat y=g(\tilde y)$ transfer \thetag{3.53} into the solution of
the equations analogous to \thetag{3.52} in new variables. Under
this change of variable the functions $\hat p$ and $\hat q$ are
transformed according to the rule
$$
\xalignat 2
&\tilde p(\tilde x)=\hat p(f(\tilde y)),
&&\tilde q(\tilde x)=\hat q(g(\tilde y)).
\tag3.54
\endxalignat
$$
The rule of transformation \thetag{3.54} let's us choose the
solutions of the equations \thetag{3.17} so that in the
appropriate variables $\tilde x$ and $\tilde y$ they are
$$
\xalignat 2
&\hat u=-K\,(\tilde x+\tilde y),
&&\hat v=-L\,(\tilde x+\tilde y),
\tag3.55
\endxalignat
$$
i.e. $\tilde p(\tilde x)=\tilde x$ and $\tilde q(\tilde y)=
\tilde y$.  Now we are only to substitute \thetag{3.55} into
the formulas \thetag{3.20}. For $\tilde F$ and $\tilde B$
such substitution gives
$$
\xalignat 2
&\tilde F=\frac{3}{K\,L\,(\tilde x+\tilde y)^2},
&&\tilde B=-\frac{9}{K\,L^2\,(\tilde x+\tilde y)^3}.
\tag3.56
\endxalignat
$$
Further we should find $\tilde P$ and $\tilde S$. This is
also done by means of relationships \thetag{3.20}
$$
\xalignat 2
&\tilde P=-\frac{L}{3\,K^2\,(\tilde x+\tilde y)},
&&\tilde Q=-\frac{1}{\tilde x+\tilde y}.
\tag3.57
\endxalignat
$$
In order to calculate $\tilde R$ and $\tilde S$ we shall
use the formulas \thetag{3.21}. From them we derive
$$
\xalignat 2
&\tilde R=-\frac{-3\,I_3-2}{3\,(\tilde x+\tilde y)},
&&\tilde S=-\frac{I_7\,K}{L^2(\tilde x+\tilde y)}.
\tag3.58
\endxalignat
$$
By substituting \thetag{3.57} and \thetag{3.58} into \thetag{2.19}
we find the relationships that bind $I_3$ and $I_7$ with $K$ and $L$
$$
\aligned
&3\,K\,I_7+3\,(K^2-2\,L)\,I_3+6\,L^2-8\,L\,K^2=0,\\
\vspace{1ex}
&6\,K\,I_3^2-7\,K\,L\,I_3-8\,K^2\,I_7-I_7\,L+9=0.
\endaligned
\tag3.59
$$
Formulas \thetag{3.57}, \thetag{3.58} and \thetag{3.58}
completely determine the canonical form of the equation
\thetag{1.1} with identically constant invariants when $F\neq 0$,
$K\neq 0$ and $L\neq 0$.
\head
4. Final remarks and acknowledgments.
\endhead
    Problem of equivalence considered in this paper has
a long history started from the last century (see \cite{10}).
Here are some references to the papers concerning this problem.
References \cite{11} and \cite{12} are the papers by E.~Cartan.
Papers \cite{2}, \cite{11} and \cite{12} are translated into
Russian and published in the book \cite{13}. We are grateful
to E.G.~Neufeld, who gave us to read this book. References
\cite{6}, \cite{14--17} are communicated us by V.V.~Sokolov
and V.E.~Adler. \par
    We are grateful to E.G.~Neufeld, V.V.~Sokolov, V.E.~Adler
and N.~Kamran for the information and helpful
instructions.

\enddocument
\end